\documentclass[a4paper, 11pt]{article}
\usepackage[utf8]{inputenc}
\usepackage{graphicx}
\usepackage{natbib}
\usepackage{comment}
\usepackage{authblk}

\usepackage{lipsum}

\title{Interaction of a migrating cell monolayer with a flexible fiber}

\author[1,2]{Leticia Valencia}
\author[3]{Ver\'onica L\'opez-Llorente}
\author[5]{Juan C. Lasheras}
\author[1,2,3]{Jos\'e L. Jorcano}
\author[2,4]{Javier Rodr\'{\i}guez-Rodr\'{\i}guez\footnote{bubbles@ing.uc3m.es}}

\affil[1]{Dept. of Biomedical and Aerospace Engineering, Carlos III University of Madrid, Spain}
\affil[2]{Academic Unit for Disruptive Technologies in Regenerative Medicine\\

Carlos III University of Madrid, Spain}
\affil[3]{Division of Epithelial Biomedicine, CIEMAT-CIBERER, Madrid, Spain }
\affil[4]{ Department of Thermal and Fluid Engineering,
 Carlos III University of Madrid, Spain}
\affil[5]{Department of Mechanical and Aerospace Engineering and\\

Department of Bioengineering, University of California San Diego, La Jolla, USA}

\date{}

\begin{document}

\maketitle

\begin{abstract}
Mechanical forces influence the development and behavior of biological tissues. In many situations these forces are exerted or resisted by elastic compliant structures such as the own-tissue cellular matrix or other surrounding tissues. This kind of tissue-elastic body interactions are also at the core of many state-of-the-art {\it in situ} force measurement techniques employed in biophysics. This creates the need to model tissue interaction with the surrounding elastic bodies that exert these forces, raising the question: which are the minimum ingredients needed to describe such interactions? We conduct experiments where migrating cell monolayers push on carbon fibers as a model problem. Although the migrating tissue is able to bend the fiber for some time, it eventually recoils before coming to a stop. This stop occurs when cells have performed a fixed mechanical work on the fiber, regardless of its stiffness. Based on these observations we develop a minimal active-fluid model that reproduces the experiments and predicts quantitatively relevant features of the system. This minimal model points out the essential ingredients needed to describe tissue-elastic solid interactions: an effective inertia and viscous stresses.
\end{abstract}

\section*{Introduction}

The forces applied by or on a living tissue have a strong impact on its behavior and development. During embryogenesis, the forces exerted by individual cells influence differentiation, migration and proliferation at the tissue level, which in turn contributes to shape the future organism \citep{DAngelo_etalCurrBiol2019}. Forces also regulate wound healing. For instance, in skin and cornea, healing was observed to occur by a  migrating tongue-shaped epithelium. These tongue-shaped projections develop from either side of the remaining intact epithelium and advance until they meet \citep{DuaForresterAmJOph1990, WANG20161443}. In general, in wound repair, cell monolayers migrate until they encounters another tissue that halts their motion by applying a pressure, which is known as contact inhibition \citep{MolinieGautreau_MethodsMolBiol2018}. Forces are important not only in the development of healthy tissue, but also in the origin and progression of some diseases, most notably cancer \citep{Foolen_etalJPD2016, Dolega_etalNatComm2017}.
For example, it has been recently reported that exerting compression stresses on a tumor makes it more resistant to chemotherapeutic drugs. This effect is mediated by the negative effect that compressing the tumor has on cell proliferation \citep{Rizzuti_etalPRL2020}.
All these situations have in common that a cell tissue exchanges forces with external constraints whose mechanical properties in turn affect the tissue's behavior. One of the key questions to understand this kind of interactions is whether and how a tissue modulates the force it can exert depending on the compliance of its surroundings. Answering this question in a model system, a migrating epithelial cell monolayer, is the first motivation of this paper. We have chosen this model system because, despite its simplicity, it is representative of what happens in wound healing situations \citep{Safferling_etalJCB2012, DuaForresterAmJOph1990, WANG20161443}.

The importance of mechanical forces in the evolution of biological tissues has motivated the development of a variety of measurement techniques able to quantify these mechanical stresses \citep{Sugimura_etalDevelopment2016, PolacheckChenNatMeth2016, Roca-CusachsNCB2017, Zhang_etalFBB2019}.
Generally speaking, these techniques rely on the deformation of an elastic body of known properties. For instance, \citet{Campas_etalNatMethods2014} proposed to use a drop embedded in a three-dimensional tissue. Different authors have refined this procedure and made it more reproducible by replacing the drop by elastic beads \citep{Dolega_etalNatComm2017, Mohagheghian_etalNatComm2018, Girardo_etalJCB2018}. Moreover, beads can be used not only to probe forces, but also to exert them, which allows the direct quantification of the tissue's mechanical properties \citep{DAngelo_etalCurrBiol2019}.\\

Although force measurements are informative by themselves, their physical and biological interpretation benefits from theoretical mechanical models which, in turn, feed on experimental measurements for their validation and tuning. Many of these models describe the tissue as fluid- or solid-like active-matter characterized by fields such as velocity, normal and shear stress, or cell density \citep{AlertTrepatARCMP2019, BanerjeeMarchetti2019, Moitrier_etalSoftMatter2019}. These quantities obey conservation laws which are expressed as partial differential equations.

To reduce the complexity of the models, an important question is: which are the essential effects that need to be modeled to describe a certain behavior?
For instance, phenomena as complex as the appearance of mechanical waves \citep{Serra-Picalmal_etalNatPhys2012} have been explained by continuum models that take into account the exchange between elastic energy and cell polarity \citep{Tlili_etalRSOS2018}, which can be interpreted in this context as a gradient in the concentration of contractile molecules \citep{Banerjee_etalPRL2015, ValenciaPhD2017}. These models are not only useful to qualitatively describe tissue behavior. The parameters they use to fit experimental observations constitute useful metrics to compare the mechanical properties of different tissues, which in turn is useful to characterize the effect of applying genetic or pharmacological treatments to the cells. A good example of this are the so-called competition assays, where two confronting cell populations migrate in opposite directions until they meet. By fitting the observed velocity and stress fields with fluid-like behaviors, Moitrier and co-workers were able to infer the mechanisms that a mutated cell population uses to push back a non-mutated one \citep{Moitrier_etalSoftMatter2019}. We highlight here that, for these metrics to be practical, they must be as reduced as possible: the lesser parameters needed to describe tissue behavior, the easier their biological interpretation will be. In this work, we keep this idea as a guiding principle when deriving our theoretical model.

In the examples given above there exist relatively simple mathematical models that describe rather complex behaviors observed in tissues free of external stress. Notwithstanding, there are no models that describe how the force that a tissue is able to exert is modulated by the compliance of an external constraint. The closest experimental systems reported in the literature are the so-called competition assays mentioned before. However, although mechanical forces clearly influence the outcome of these experiments, there are also complex biochemical interactions taking place at the interface when the two cell populations meet \citep{Moitrier_etalSoftMatter2019}. This makes hard to isolate the role of mechanics.

\begin{figure}[hbt!]
\centering
\includegraphics[width=0.6\columnwidth]{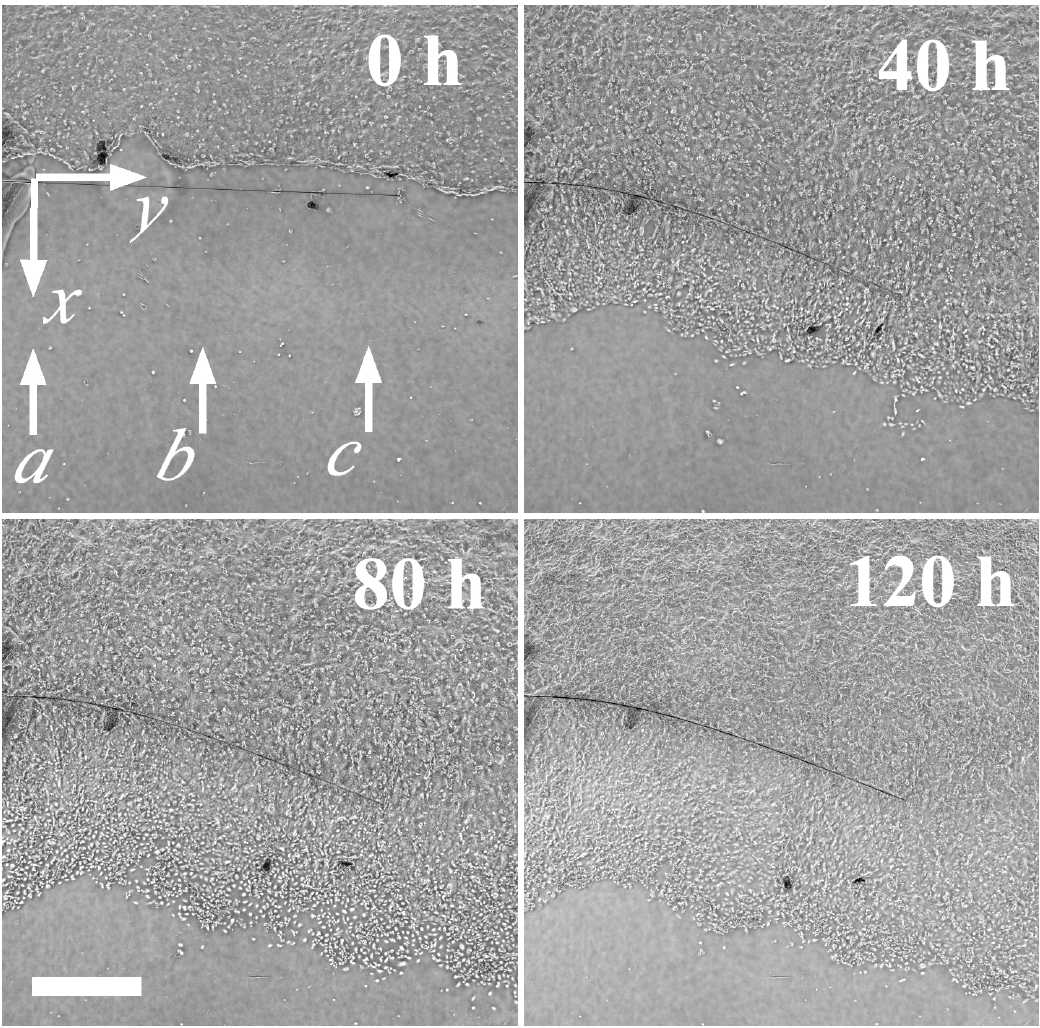}
\caption{\label{fig:mosaic-time} Time lapse showing the progress of a migrating monolayer interacting with a flexible carbon fiber. The $x$ axis denotes the direction of migration, whereas the $y$ axis runs parallel to the initial, undeformed, fiber. ($a$),($b$) and ($c$) denote the three regions used to obtain the kymographs shown in Figure ~\ref{fig:V3mm} and Figure ~6 of the Supplemental Material. Scale bar: 1000 ${\mu}m$.}
\end{figure}

With this motivation in mind, we propose a relatively simple, and thus easy to control, configuration: a migrating epithelial cell monolayer pushing a slender millimetric-long carbon fiber fixed at one of their ends (see Figure ~\ref{fig:mosaic-time}). Despite the relative simplicity of the system, a migrating cell monolayer is commonly used in biophysics as a model problem for more complicated phenomena taking place in wound healing \citep{Safferling_etalJCB2012} or embryogenesis. Moreover, besides the fundamental interest of these experiments to aid the development of theoretical active matter models of biological tissues, the configuration we choose has potential applications in the development of an inexpensive technique that can be used to probe the force exerted by a migrating three-dimensional tissue, as will be discussed below. 

The second motivation of this paper is precisely to develop this technique in a relatively simple and widely-studied two-dimensional system. Despite its simplicity, the capacity to measure the force exerted by a tissue monoloayer on a flexible obstacle --a carbon fiber in our case-- will undoubtedly facilitate identifying and quantifying the  biophysical mechanisms involved in healthy tissues and diseases such as, among others, chronic skin ulcers and inherited skin fragility disorders (Epidermolysis bullosa), conditions on which our group has been investigating for a long time \citep{Garcia_etMolCarcinog2007}.

\section*{Materials and Methods}

\subsection*{Cell culture:} 
The human skin keratinocyte cell line (HaCaT) was  cultured in Dulbeco’s Modified Eagles Medium (DMEM, Invitrogen Life Technologies) supplemented with 10$\%$ Fetal Bovine Serum (FBS, Thermo Scientific HyClone) and 1$\%$ of Antibiotic Antimycotic (ThermoFisher) following standard protocols well established in our laboratory  \citep{Paramio2001}.\\

\subsection*{Experimental procedure:} 
\begin{figure}[hbt!]
    \centering
    \includegraphics[width=0.4\columnwidth]{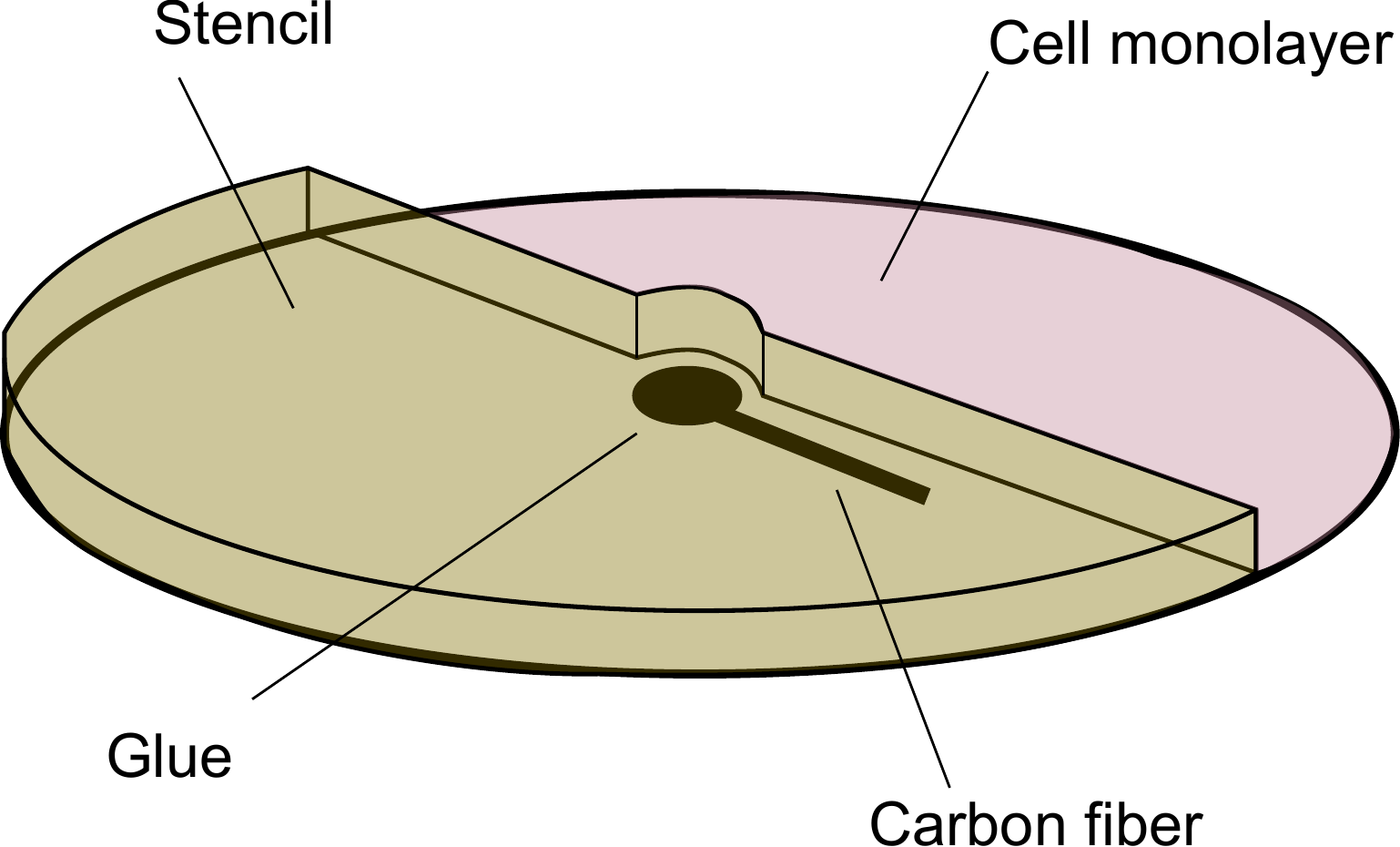}
    \caption{Sketch of the experimental set-up (not to scale). A 3D-printed stencil is used to make a repeatable wound which is always placed at a similar initial distance from the fiber. The drop of glue is outside the displayed pictures, immediately at the left of the carbon fiber.}
    \label{fig:sketch1}
\end{figure}

A short piece of a carbon fiber is manually positioned  near the center of a 35 mm cell culture Petri-dish  (Corning) and glued at one end with a drop of commercial cyanocrylate glue. The drop has to be as small as possible to minimize its interaction with the cell culture. Then, the fiber is cut to the corresponding length (2 to 5 mm) using a scalpel. The surface of the culture plate and the glued filament, are then functionalized with a solution of bovine skin type I collagen at 0.1 \% (w/v) for 2 hours at room temperature under UV light. Afterwards, a 3D printed polylactic acid (PLA) stencil is placed in the culture plate as shown in Figure ~\ref{fig:sketch1}.

Cells are then seeded at a final density of 1550 cells/mm$^2$ on the free half of the dish and incubated for 24 hours to allow them to attach to the substrate, forming a confluent monolayer. Then, the stencil is carefully removed creating the two-dimensional wound. The culture was afterwards washed with PBS, and then 4 ml of fresh culture medium is added.
This stencil-based method is more reproducible and causes less harm to the monolayer front than the classical method to scratch the tissue with a pipette tip.
\citep{Jonkman2014}.\\

Experiments are performed in an automated inverted microscope Leica Dmi8 equipped with an OKOLab incubator. A four-Petri-dish adaptor allows to control the correct temperature, air/CO2 and humidity control during the experiment \citep{Goldman2005}. The time-step between frames is 15 minutes and the total duration of each experiment is 5 days (120 hours).
Images are acquired in phase contrast with a 5X magnification objective and a Hamamatsu sCMOS Orca Flash 4.0 LT camera by means of LASX Navigator acquisition software from Leica Microsystems.\\

\subsection*{Image Analysis:}
Custom-made image analysis codes for fiber detection were developed in Matlab software, using segmentation algorithms described in \citep{Imseg2006}. The spatial resolution of the detection is 1.29 $\mu m$.\\


\subsection*{Velocity measurements:}

Velocity fields were measure from the time-lapse images using the open-source toolbox PIVLab (Time-Resolved Digital Image Velocimetry Tool for MATLAB) \citep{PIVlab}.
We set the size of the interrogation window to 128 pixels with an overlap of 64 pixels and a second pass of 32 pixels, leading to a 32-pixel step between vectors. This corresponds to a spatial resolution of approximately 42 $\mu$m between velocity vectors. The correlation algorithm chosen for the calculations was Fast Fourier Transform with multiple passes and allowing window deformation. The toolbox includes data validation section to filter noisy vectors by interpolating them between neighboring ones.\\

Kymographs are space–time plots which display intensity values of a third variable (in our case the streamwise velocity, $u$, that is the velocity along the $x$ direction in Figure ~\ref{fig:mosaic-time}) thus reducing by projection three-dimensional data ($x$, $t$, $u$) to two dimensions \citep{NITZSCHE2010247}. For each region (denoted by arrows $a$, $b$ and $c$ in Figure ~\ref{fig:mosaic-time}) three columns of PIV velocity boxes of streamwise velocity ($u$) are averaged over the spanwise ($y$) direction in a band with a width of 126 $\mu m$. Each one of these averages is represented as a function of time with the color showing the value of $u$.

\section*{Results and Discussion}

\subsection*{Experimental measurement of the force exerted by the migrating monolayer }
We conduct experiments in which a monolayer of skin epithelial cells migrates to close a wound, which we produce artificially by using a stencil to prevent cells from attaching on one half the Petri dish, as explained in Materials and Methods (see Fig. \ref{fig:sketch1}). At a fixed distance from the stencil, we glue one of the ends of a carbon fiber in such a way that it lays in cantilever parallel to the initial front of the monolayer. In all the experiments reported here, the only parameter that is varied is the length of the fiber, which effectively determines its bending stiffness. The diameter of the fiber, $d = 7.8$ $\mu$m, is comparable to the typical size of the cells, which is about 10-30 $\mu$m.

\begin{figure}[hbt!]
\centering
\includegraphics[width=\columnwidth]{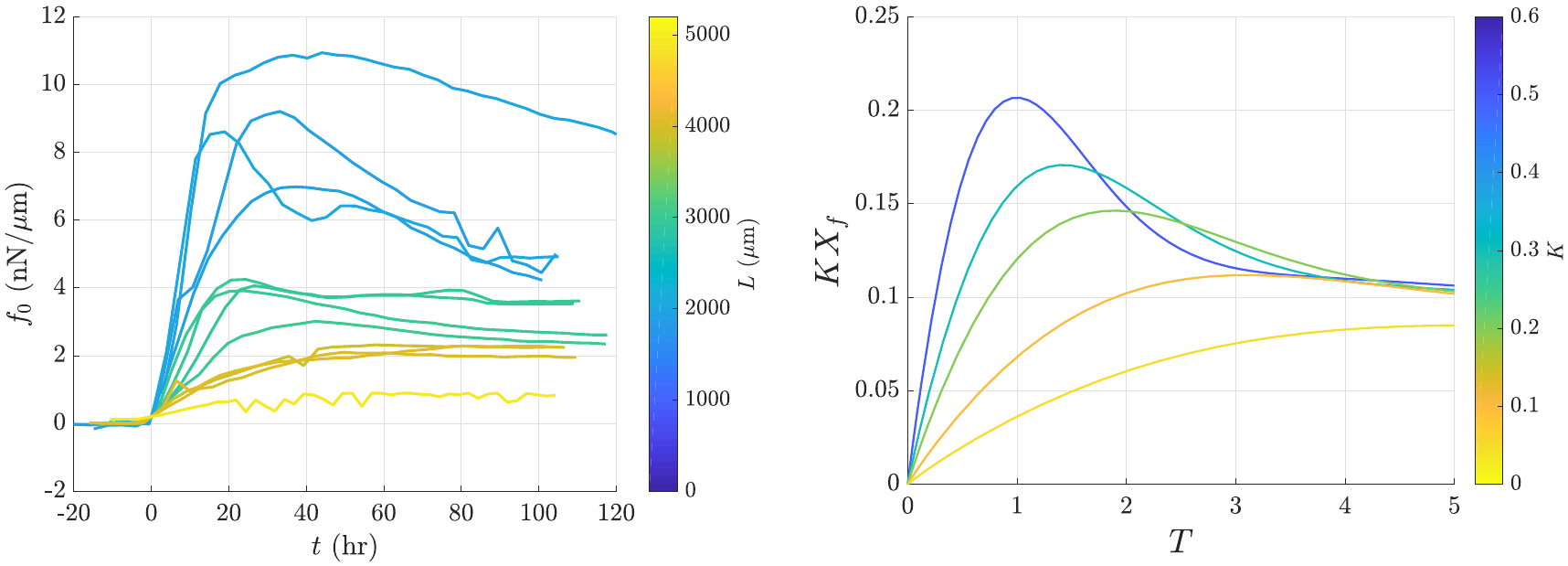}
\caption{\label{fig:forces_all_sessions} ($a$) Time evolution of the uniform force per unit length $f_0$ for different fiber lengths. The color scale represents the length of the fiber, $L$. The initial time $t=0$ is that at which the monolayer reaches the fiber. ($b$) Time evolution of the characteristic deflection of the fiber times stiffness (equivalent to a dimensionless force per unit length) computed numerically for $\hat{\ell} = 5$ and $K = 0.5, 0.3, 0.2, 0.1$ and $0.05$. Darker colors correspond in both plots to stiffer fibers.}
\end{figure}

We model the effect of the cell monolayer pushing the fiber as a uniform force per unit length, $f_0$. To compute this force we fit the equation describing the shape of a flexible fiber of length $L$, fixed at one end ($y=0$) and free at the other ($y=L$) and deflecting under a constant force \citep{Timoshenko1940},
\begin{equation}
    x = \frac{f_0}{24EI}y^2\left(y^2 - 4Ly + 6L^2\right),
    \label{eq:rod_shape}
\end{equation}
to the different shapes adopted by the fiber over time. In this work $x$ will denote the streamwise coordinate, the direction along which cells move, perpendicular to the undeformed fiber, whereas $y$ will be the spanwise coordinate, see Fig. \ref{fig:mosaic-time}. Moreover, $E$ is the Young's modulus and $I$ the moment of inertia of the fiber (for our fibers, $E = 229\times10^6$ nN/$\mu$m$^2$ and $I = (\pi/4)(d/2)^2 = 181.7$ $\mu$m$^4$). The force $f_0$ measured in this way for the different experiments is shown in Fig. ~\ref{fig:forces_all_sessions}a as a function of time.

\begin{figure}[hbt!]
\centering
\includegraphics[width= \columnwidth]{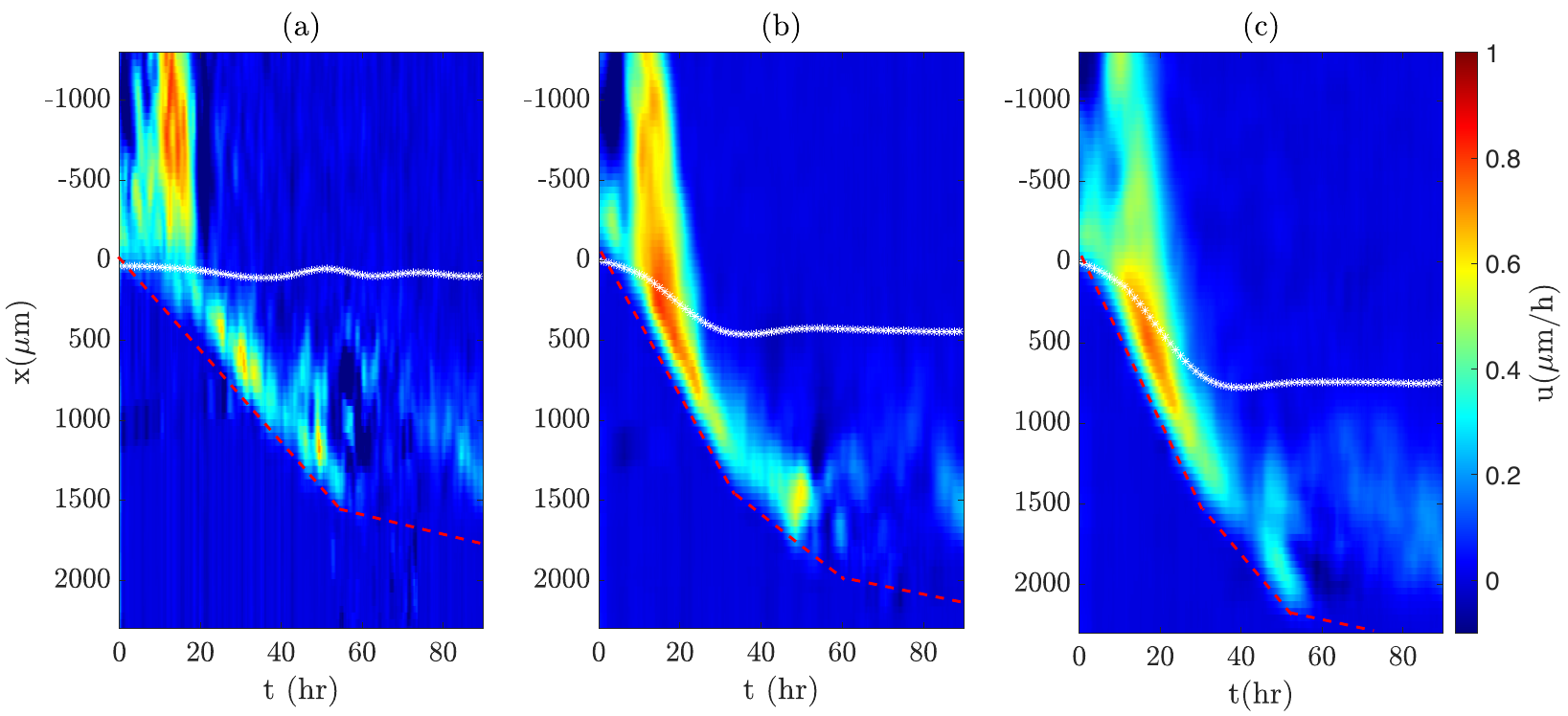}
\caption{\label{fig:V3mm} Time evolution of the velocity field $u(x, t)$ (colormap) and fiber deflection (white thick line) at three different locations, shown in Figure ~\ref{fig:mosaic-time}. The red dashed lines marks the location of the foremost cell that has managed to migrate downstream the fiber. Positive velocities are directed downwards, coinciding with the direction of the migration. Position (a), close  to the glue drop, does not experiment any deformation. Length of the fiber: $L = 3$ mm. See Supplemental Material for more kymographs.}
\end{figure}

The most salient feature of these curves is the existence of a maximum force, and thus deflection, after which the fiber recoils. Concomitantly, the monolayer stops at this point. This maximum is very prominent in the shorter fibers and is barely noticeable in the longer ones, its value not being very different from the asymptotic deflection reached at long times. The recoil of the fiber and the decay of the force start to take place when the cell monolayer stops, as can be inferred from kymographs like the one shown in Fig.~\ref{fig:V3mm}. These kymographs display the time evolution of the magnitude of the streamwise velocity, $u(x, t)$, (color scale) at the three different vertical lines indicated in Fig.~\ref{fig:mosaic-time} as $a$, $b$ and $c$. To correlate the time evolution of the velocity with the fiber deflection, we show the position of the fiber at each location with a thick white line. The dashed red line denotes the time evolution of the foremost cells. As is also apparent in Fig. ~\ref{fig:mosaic-time}, some cells migrate downstream the fiber after the monolayer touches it. These cells continue migrating even after the fiber has stopped. However, the amount of cells that overpass the fiber is small compared with those that remain pushing it upstream. As a consequence, the cell density downstream the fiber is substantially smaller than that upstream (see Supplemental Material, Fig.~9). This means that the percentage of cells surpassing the fiber is small and, consequently, that the force with which they may pull from it must be small compared to the push exerted by the bulk monolayer. At the same time, because few cells pass beyond the fiber, we do not expect their passage to diminish the push exerted by the bulk.

\subsection*{Model formulation }
Modelling in full the motion of the cell monolayer is a very complicated task that lies beyond the scope of this work. Instead, we propose here a minimal model that captures the physics of the interaction of the cell monolayer with a flexible fiber that opposes its motions. The one-dimensional model proposed here treats the cell monolayer as a compressible active fluid with velocity field $u(x, t)$. The model incorporates the effect of an effective inertia, Newtonian viscous stresses and a Hookean linear force exerted by the fiber on the tissue. The choice of these effects is motivated by the experimental results discussed below. Under these assumptions, momentum conservation reads
\begin{equation}
    m\left(\frac{\partial u}{\partial t} + u\frac{\partial u}{\partial x}\right) = -k x_f \delta\left[x-x_f\right] + \mu \frac{\partial^2 u}{\partial x^2}.
    \label{eq:navier-stokes}
\end{equation}
The left-hand side of the equation, equivalent to an inertia, accounts for the resistance of a cell to change its state of motion \citep{Banerjee_etalPRL2015}. The coefficient $m$ regulates the importance of this effect. Note that this effective inertia has a very different origin from the real (i.e. mechanical) one, which is always negligible in cell mechanics \citep{AlertTrepatARCMP2019}. 
In fact, borrowing ideas from the model by Blanch-Mercader and co-workers \citep{BlanchMercader_etalSoftMatter2017}, we could associate the effective inertia $m$ to the time it takes for cells to change their polarization (an internal property that modulates the speed and direction of their migration) in response to an external force. Although these authors assume that in a freely migrating monolayer the relaxation time of the polarization field is much shorter than any other time scale of the system, in our problem cells encounter the fiber suddenly, so it is reasonable to assume that the relaxation dynamics of the cell polarization field manifest through an inertia-like term. Such a term is neglected in Equation (4) of reference \citep{BlanchMercader_etalSoftMatter2017}. We must also point out that we have used the material derivative $D/Dt = \partial /\partial t + u \, \partial / \partial x$, since it represents the time derivative of a flow property computed following a migrated cell \citep{Batchelor1967}.

The first term in the right-hand side, the elastic force exerted by the fiber is proportional to a characteristic deflection, $x_f$, times a proportionality constant $k$, the bending stiffness. This deflection could be, for instance, that at the tip. From equation (\ref{eq:rod_shape}) it is possible to infer that $k = \partial f_0/\partial x_f \sim EI/L^4$.

Finally, the last term on the right-hand side accounts for viscous stresses. These arise from the friction exerted between moving cells \citep{AlertTrepatARCMP2019}. The way we model viscous forces in our monolayer is identical to that of Blanch-Mercader and co-authors \citep{BlanchMercader_etalSoftMatter2017}.

Equation (\ref{eq:navier-stokes}) needs two boundary conditions. One is going to be imposed a distance $\ell$ upstream of the initial position of the fiber, i.e. at $x = -\ell$. It is reasonable to assume that, sufficiently far from the fiber, the velocity of the cells does not any longer depend on the position, thus $\partial u/\partial x = 0$ there. The other boundary condition is applied at the front of the monolayer, which coincides with the position of the fiber $x = x_f(t)$. This position needs to be computed as part of the solution. To deal with this moving boundary it is convenient to introduce a scaled spatial coordinate, $\xi = \left(x - x_f\right)/\left(\ell + x_f\right)$, such that $\xi(x=-\ell) = -1$ and $\xi(x=x_f)=0$. Introducing this new variable turns equation (\ref{eq:navier-stokes}) into:
\begin{equation}
    \frac{\partial U}{\partial T} + \frac{U-\dot{X}_f}{\hat{\ell} + X_f}\frac{\partial U}{\partial\xi} = -K X_f \delta\left[\xi\right] + \frac{1}{\left(\hat{\ell} + X_f\right)^2}\frac{\partial^2 U}{\partial\xi^2}.
    \label{eq:navier-stokes_rescaled}
\end{equation}
We have introduced the following dimensionless notation: $U = u/u_0$, $T = t u_0 / L_c$, $\hat{\ell} = \ell/L_c$, $K = kL_c^3/\mu u_0$ and $X_f = x_f/L_c$. Here $u_0$ is the velocity of the monolayer at the time it touches the fiber --which we assume uniform, at least in a region of size $\ell$-- and $L_c = \mu/m u_0$. This length scale $L_c$ measures the size of the region where the effective inertia is of the order of the viscous stresses. As will be discussed later in view of the experimental results, we expect this to be the characteristic size of the flow.

The boundary condition at the front can be found by integrating equation (\ref{eq:navier-stokes_rescaled}) along an infinitesimal interval centered at $\xi=0$. Doing so, we get
\begin{equation}
    \frac{\dot{X}_f^2}{2\left(\hat{\ell} + X_f\right)} + K X_f + \frac{1}{\left(\hat{\ell} + X_f\right)^2}\left.\frac{\partial U}{\partial\xi}\right|_{\xi=0} = 0.
\end{equation}
Finally, as in any free-boundary problem, the above equations need to be completed with the kinematic boundary condition imposing that the boundary moves with the local velocity
\begin{equation}
    \dot{X}_f = U(\xi=0).
\end{equation}

Note that the boundary condition imposed at the front of the migrating monolayer amounts, in fact, to assume that cells do not overpass it.

\subsection*{Comparison between experimental and model results}

The fiber-monolayer system behaves as a damped harmonic oscillator, close to or around the critical damping, endowed with an initial kinetic energy. Initially, the fiber deflects at a nearly constant speed, progressively slowing down until it reaches a maximum deflection. The stiffer (shorter) fibers investigated here slowly recoil after reaching the maximum deformation, whereas the less stiff (longer) ones asymptotically approach the maximum deformation and keep it for as long as we can observe (see Figure ~\ref{fig:forces_all_sessions}a).

These observations suggest the choice of two fluid-like behaviors to build a minimal model: a) an inertia-like term, as the tissue does not immediately modify its velocity upon touching the fiber; and b) a viscous damping to account for the slow recoil and the fact that at most only one relative maximum deflection is reached.

The minimal model built by balancing these two effects with the elastic recovery force of the fiber captures qualitatively the experimental observations. To illustrate this, Figure ~\ref{fig:forces_all_sessions}b portraits the time evolution of the fiber deflection times the stiffness, $K X_f$, predicted by the model for different stiffness.

The fact that the fiber deflection exhibits overdamped or nearly overdamped oscillations points out that inertia and viscosity are equally important in the motion of the system. Establishing an analogy with the flow of a viscous fluid, we could say that the monolayer motion has a Reynolds number, $Re = m u_0 L_c / \mu$, of order unity. Consequently, if $Re \sim 1$, then $L_c = \mu / m u_0$ can be chosen as the length scale of the flow.
At long times, the model predicts that the whole monolayer stops and the fiber recoils back to equilibrium. Note that, in our experimental setup, it is not possible to observe times longer than about 120 hours (5 days). Thus, we cannot confirm whether the fiber goes back to its original position. The time scale over which the recoil occurs is associated to the length of tissue affected by the presence of the fiber, $\ell$. The existence of this distance where the tissue behavior is not affected by the dynamics of the front is an ingredient found in similar models of monolayers opposing some resistance, like that of \cite{Moitrier_etalSoftMatter2019}.

Our model also explains the effect of the fiber elasticity on the dynamics of the monolayer. Since the Young's modulus and cross-sectional moment of inertia of our fibers cannot be changed, we have carried out experiments with different fiber lengths, $L$ (see Figure ~\ref{fig:forces_all_sessions}a). Note that, as stated above, the stiffness of the fiber $k \sim L^{-4}$, so changing $L$ by a factor of two actually allows us to cover more than an order of magnitude in stiffness. This is supported by the variation in $K$ needed to replicate qualitatively the experimental results (see Figs.~\ref{fig:forces_all_sessions}a and b).

\begin{figure}[hbt!]
\centering
\includegraphics[width=\columnwidth]{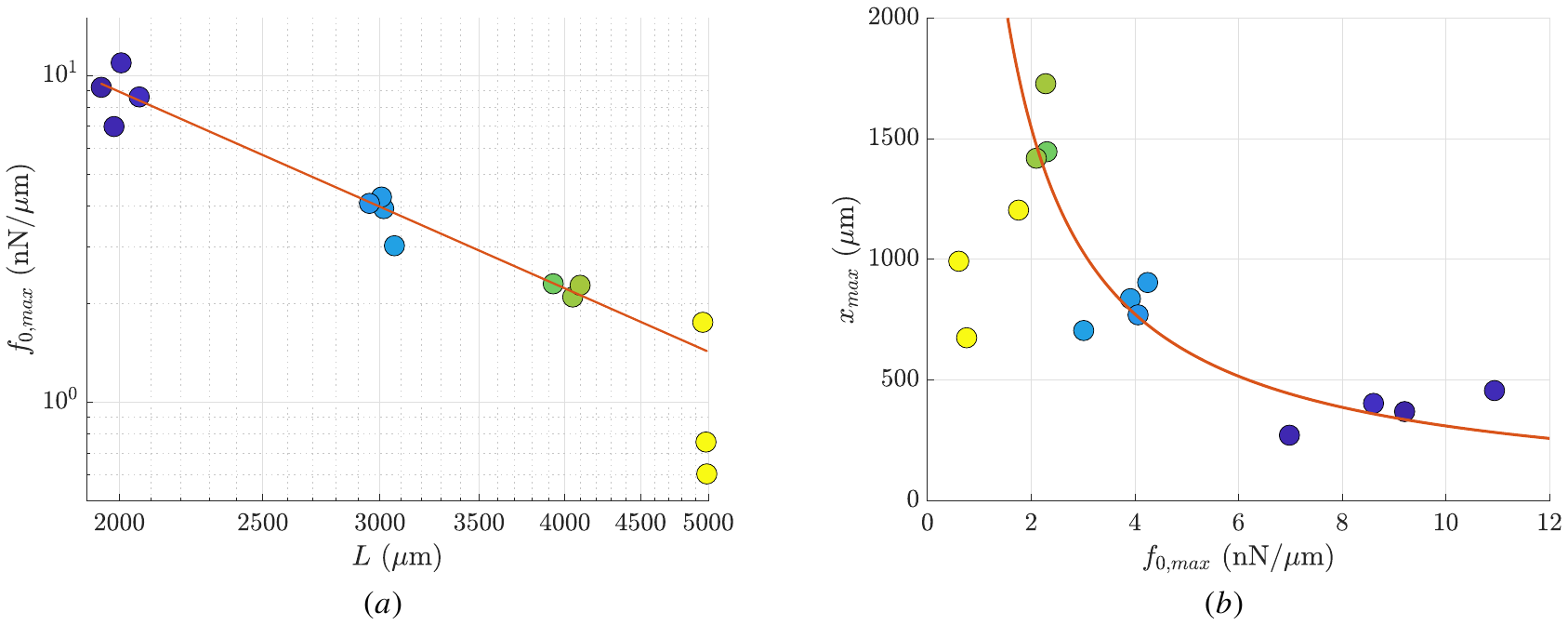}
\caption{\label{fig:force_vs_L_83slope}($a$) Maximum force measured in experiments corresponding to fibers of different lengths. Each point corresponds to an experiment with a different fiber, whose color indicates its length (see colorbar in Figure ~\ref{fig:forces_all_sessions}$a$). The solid line is a fit with a power law $f_{0, max} \sim L^{-2}$. ($b$) Maximum fiber deflection, computed as $x_{max} = f_{0, max}/k$, as a function of the maximum force. The color code is identical to that of panel ($a$). The solid line corresponds to a maximum deflection inversely proportional to the maximum force, $x_{max}f_{0,max} \approx 3.1\times10^3$ nN.}
\end{figure}

Let us assume that the cell velocities are of the order of the initial one, $u_0$, and that the maximum deflection reached by the fiber is of the order $x_{max}$. Treating the monolayer as an undamped oscillator, it is reasonable to assume that the maximum deflections reached by the fiber will take place when the elastic forces can no longer oppose inertia. Thus, balancing the elastic and inertial terms in Equation (\ref{eq:navier-stokes}) we get $x_{max}^2 \sim m u_0^2 / k$. Using Hooke's law, $f_{0,max} \sim k x_{max}$, and recalling that $k \sim EI L^{-4}$,
\begin{equation}
    f_{0,max} \sim k x_{max} \sim m^{1/2} u_0 k^{1/2} = m^{1/2} u_0 \left(EI\right)^{1/2} L^{-2}.
    \label{eq:scaling_law_f0c}
\end{equation}
This prediction is in very good agreement with the dependency of the maximum force measured for fibers of different lengths, shown in Fig. \ref{fig:force_vs_L_83slope}$a$.

A more interesting feature of the monolayer behavior follows from these arguments. The maximum elastic energy accumulated in the fiber per unit length is of the order of the product of its maximum deflection times the maximum force. Using the same order-of-magnitude estimations as above,
\begin{equation}
    E_{max} \sim f_{0,max} x_{max} \sim k x_{max}^2 \sim m u_0^2.
\end{equation}
In other words, we expect the maximum elastic energy accumulated by the fiber to be nearly independent of its stiffness. We can see in Figure ~\ref{fig:force_vs_L_83slope}$b$ how this prediction works reasonably well, except for some of the longest fibers (It seems reasonable to argue that the hypotheses of uniform force, cell velocity, etc. work worse for such long fibers, 5 mm). From the biophysical point of view, this observation puts forward a suggestive hypothesis: the cell monolayer stops its motion when the energy required to maintain a given stress overcomes that provided by cell metabollism, with little influence of the fiber stiffness.

We observe in our experiments a feature commonly found in mechanotransduction: the stronger the resistance of the environment, the stronger the force cells are able to exert \citep{Hoffman_etalNature2011}. In our model, the effective inertia makes cells push the fiber at their initial speed, until this inertia is balanced by the elastic force exerted by the fiber. Consequently, since at short times this velocity does not seem to depend on the fiber stiffness, the stiffer the fiber the stronger the force the monolayer exerts.

Besides the agreement with the model's prediction, the figure also illustrates the good quantitative repeatability of the results. The good repeatability of the results, together with the fact that the model predicts quite well the fiber deflection opens up the door to use this setup as an easy-to-implement and inexpensive tool to measure the force that a migrating tissue is able to exert in real physiological conditions. Although the experiments reported here correspond to a two-dimensional cell monolayer, it is in principle feasible to make other kinds of cells grow and develop in three dimensions embedding the fiber in the process. This is currently ongoing work and we expect to report these results in the future. We remark here that relatively few techniques exist nowadays to measure forces in three-dimensional tissues \citep{Zhang_etalFBB2019, DAngelo_etalCurrBiol2019}. In particular, our model highlights the importance of the tissue's effective inertia and viscosity in order to determine its ability to exert forces when opposed by an external body.

Finally, we expect our experimental method and the associated model to be useful to help the interpretation of the so-called confrontation assays \citep{Moitrier_etalSoftMatter2019}. In these experiments, two monolayers of cells of different nature (e.g. one of them having a mutation) migrate in opposite directions until they meet. After this happens, one of the monolayers pushes back the other. These assays allow researchers to elucidate, for instance, how a particular mutation affects the ability of a monolayer to exert forces or to invade adjacent tissues. The interpretation of the results is usually complex, as when the fronts of the monolayers come in touch, cells not only respond to mechanical forces exerted by their opponents, but also to biochemical signals triggered by the contact between their membranes. Contrarily, the experimental methodology proposed in this work assesses the response of the cells to a pure mechanical force exerted by an inert elastic body whose behavior is perfectly characterized. Moreover, not only the force can be characterized accurately: for instance, changing the functionalization of the fiber we can induce specific responses in the cells. In the experiments reported here the fiber is functionalized with collagen, so cells attach to the fiber in the same way that they attach to the substrate. All this allows our experimental methodology to be useful to identify and quantify the relative relevance of the biophysical mechanisms that affect a monolayer, either containing normal or diseased cells, which is hard or impossible to do by other methodologies.

\section*{Conclusions}

We carried out experiments where a migrating monolayer of keratinocytes pushes on a carbon fiber of known properties. Initially, the monolayer deflects the fiber at a nearly constant speed upon coming in touch with it. Later on, the velocity slows down, eventually reaching a maximum deflection after which the fiber recoils and the monolayer comes to a stop.
This behavior resembles that of a harmonic oscillator around the critical damping.

These observations inspire a minimal model that successfully describes the main features of the interaction of a migrating cell monolayer with an elastic slender fiber observed experimentally.
Although the model does not aim at describing the spatio-temporal dynamics of the tissue in its full spatial extension, it illustrates the essential ingredients needed to describe the interaction of a migrating tissue with a slender elastic object. Namely, an effective inertia, viscous stress, and an external elastic force due to the fiber.

A simple order-of-magnitude balance between cell monolayer inertia and fiber elasticity is consistent with one of the most salient features of the system: the fiber stops when cells have exerted a fixed mechanical work on it. Consistently with this, the maximum force exerted on the fiber increases with its stiffness. This puts forward a very suggestive conclusion: in the conditions under study, what limits the ability of a tissue to expand against an external constraint is not so much the force it must exert, but the associated work it needs to deliver.

Besides the purely fundamental interest of the problem in the fields of biophysics and active matter we envision two potential applications for our experimental methodology and the associated model. First,
we expect our experiments and theoretical model to pave the way for the development and improvement of experimental techniques to measure the force exerted by a three-dimensional developing tissue {\it in vivo}.

Second, our model can be used to complement other more comprehensive descriptions of migrating cell monolayers \citep{Banerjee_etalPRL2015, ValenciaPhD2017, Tlili_etalRSOS2018, Moitrier_etalSoftMatter2019} to allow them to describe the interaction with a compliant external body. For example, it could help to disentangle the roles of mechanical and biological cues in confrontation assays.

Third, even being a two-dimensional model, it can be applied to describe physiological and pathological situations such as skin and cornea wound healing, where epithelial cell monolayer migration is at the center of process.

\section*{Acknowledgments}

We thank Dr. Gustavo V\'{\i}ctor Guinea for providing us with the carbon fibers.
Javier Soler and Fernando Garc\'{\i}a collaborated in early stages. We also thank Ang\'elica Corral and Guillermo Vizca\'{\i}no for their technical assistance in the experiments. We are indebted to Dr. L. Champougny for her valuable comments on the manuscript and results.

We acknowledge the support of the Spanish Ministry of Economy and Competitiveness through grants DPI2014-61887-EXP, DIP2015-68088-P, DPI2017-88201-C3-3-R and DPI2018-102829-REDT, partly funded with European funds. Juan C. Lasheras thanks Universidad Carlos III de Madrid and Banco Santander for financial support from a Chair of Excellence.

 \bibliographystyle{biophysj}


\newpage

\section*{Supplemental Material}

\subsection*{Kymographs and velocity measurements}
\begin{figure}[hbt!]
\centering
\begin{tabular}{c}
\includegraphics[width=0.8\columnwidth, trim=60mm 5mm 30mm 3mm, clip]{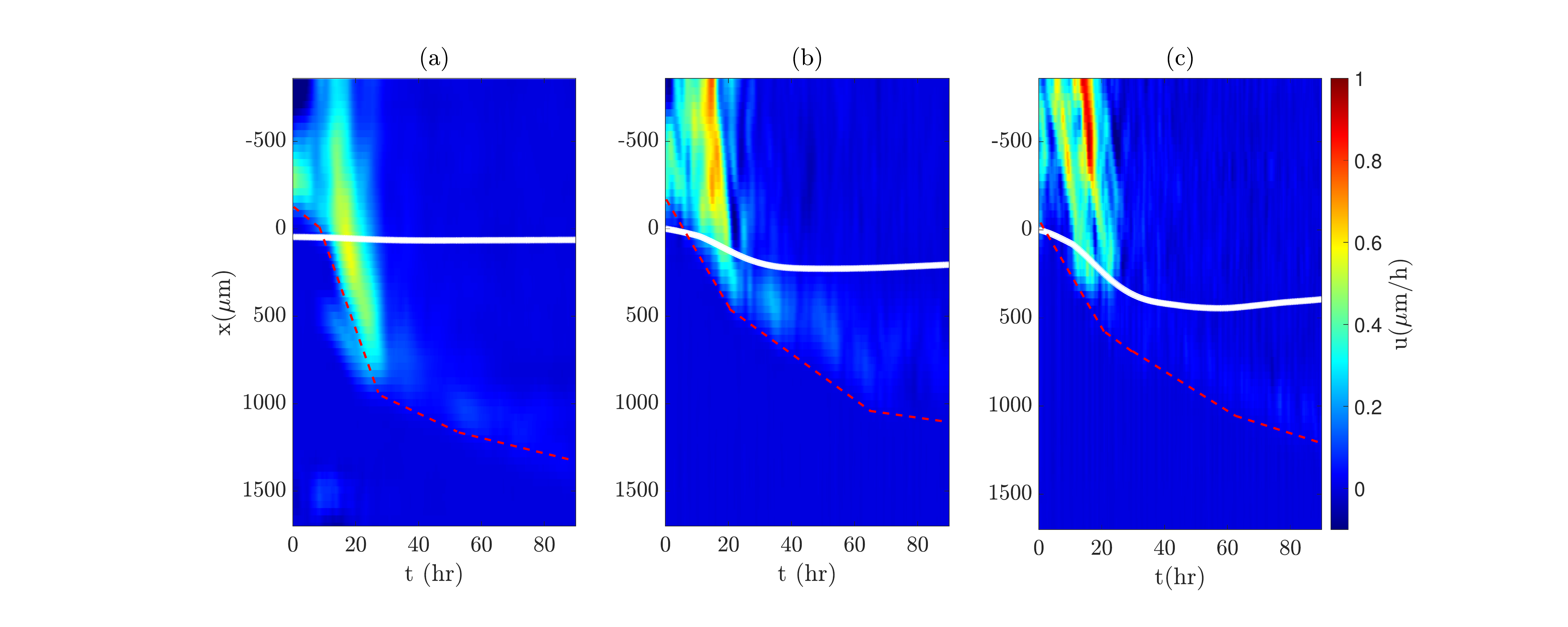} \\
\includegraphics[width=0.8\columnwidth, trim=60mm 5mm 30mm 3mm, clip]{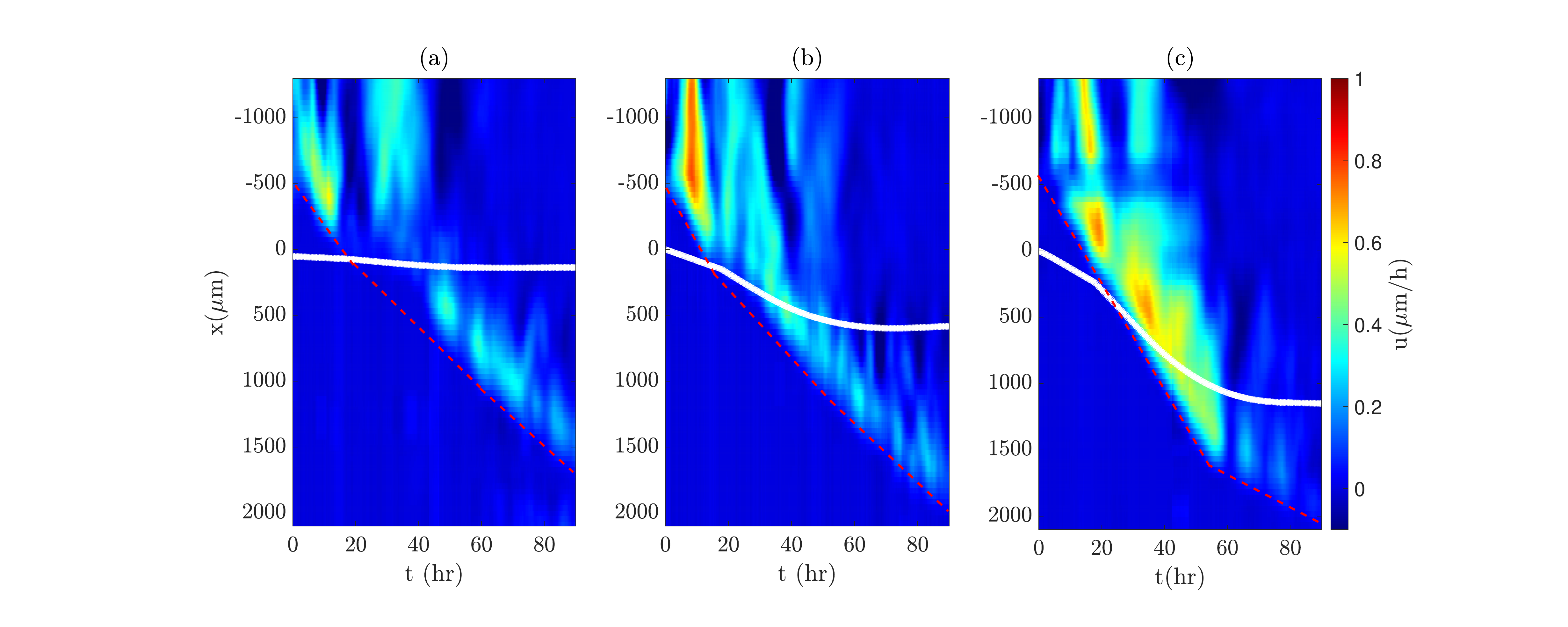}
\end{tabular}
\caption{\label{fig:V2-4mm} Time evolution of the velocity field $u(x, t)$ (colormap) and fiber deflection (white thick line) at the three different locations. ($a$), ($b$) and ($c$) correspond to the regions denoted by arrows in Figure ~\ref{fig:mosaic-time} of the main text. The red dashed lines marks the location of the foremost cell that has managed to migrate downstream the fiber. Velocities are positive when directed downwards, along the direction of the migration. Length of the fiber: $L = 2$ mm and $L = 4$ mm for the top and bottom kymographs respectively.}
\end{figure}

\begin{figure}
    \centering
    \begin{tabular}{cc}
    \includegraphics[height=50mm]{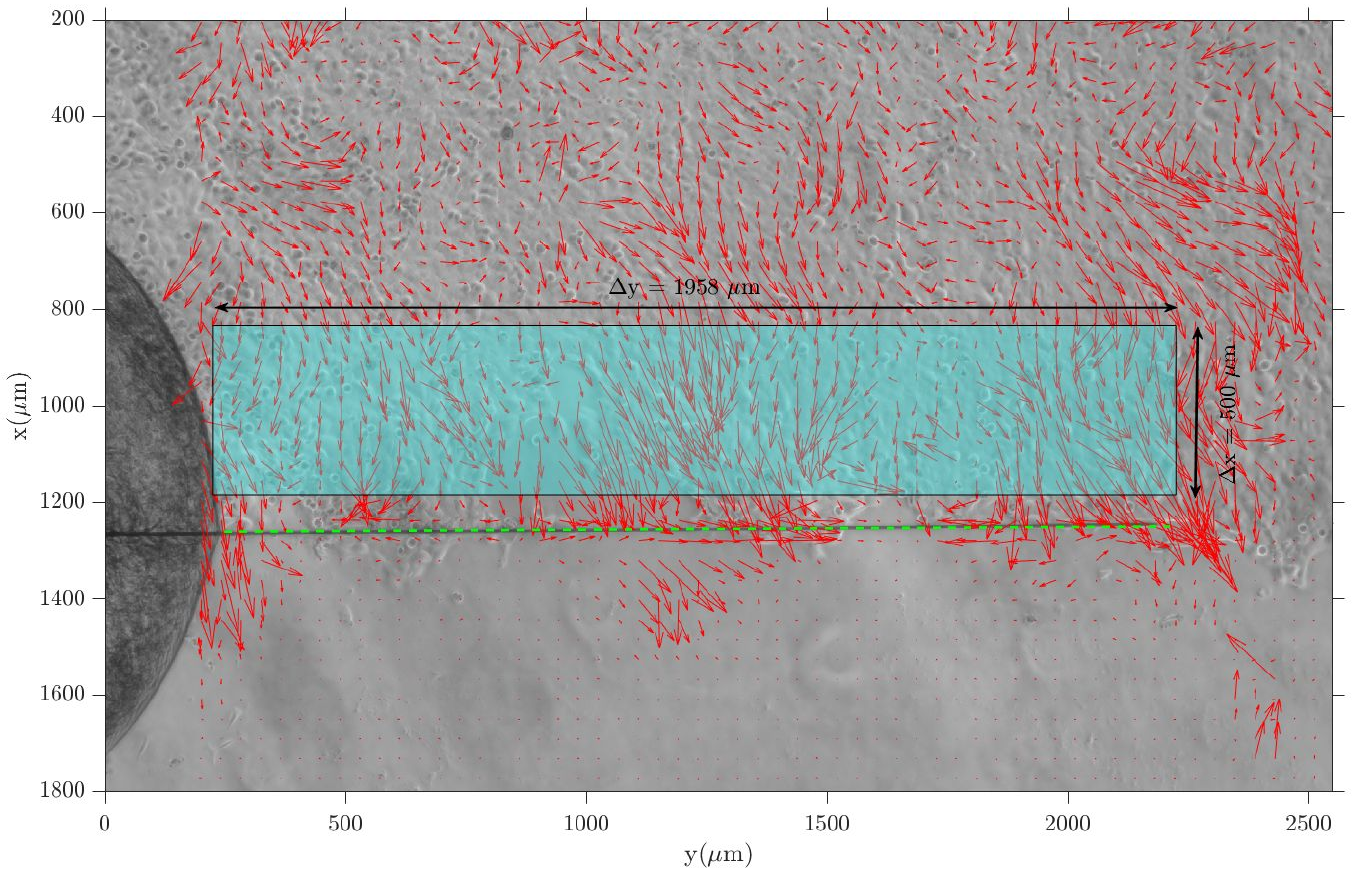} &
    \includegraphics[height=50mm]{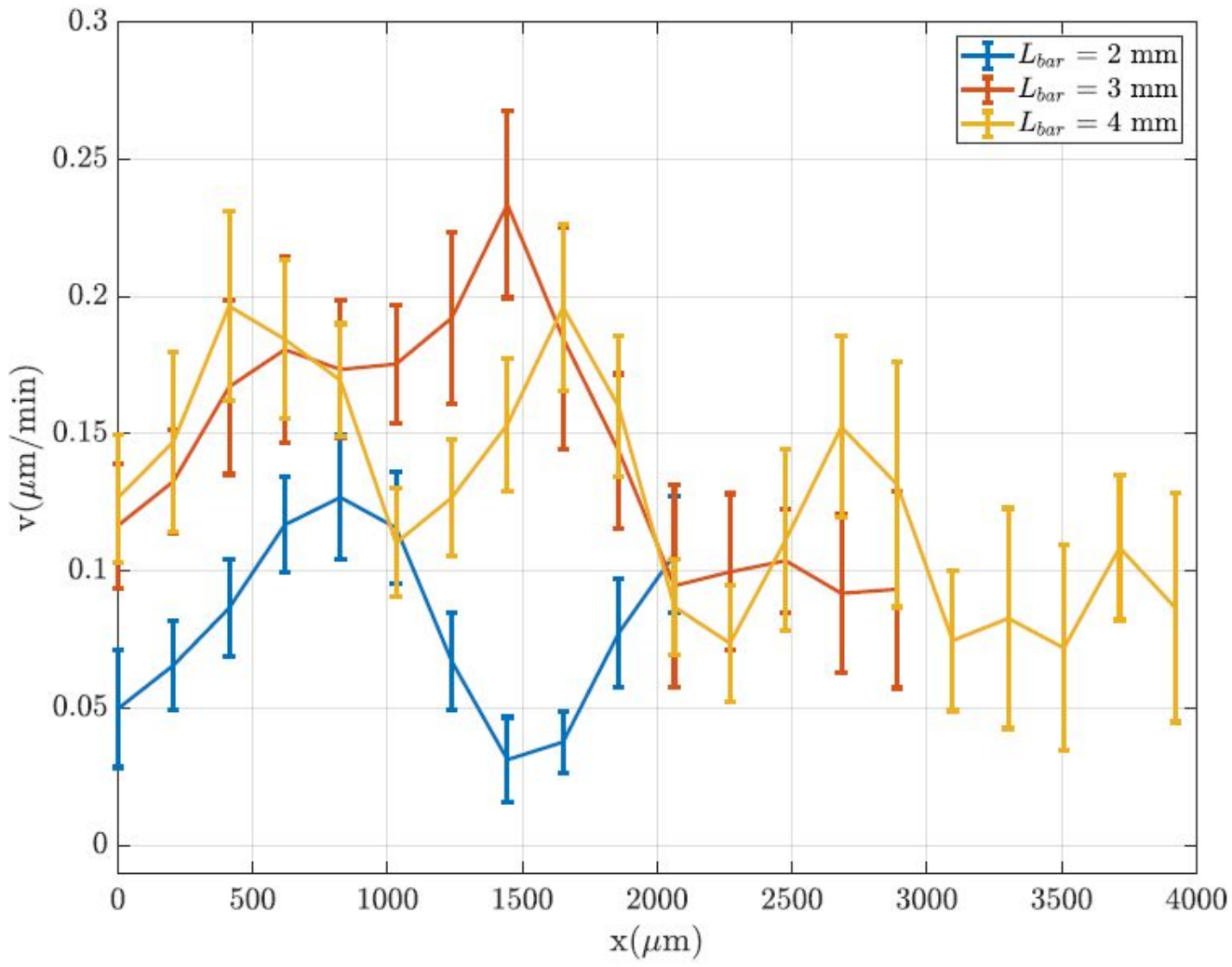} \\
    (a) & (b) \\
    \end{tabular}
    \caption{(a) Cell velocity field measured with PIV. The light blue band represents the region in which velocities are averaged over the streamwise ($x$) direction. Figure (b) represents this $x$-averaged velocity as a function of the spanwise ($y$) direction for three fibers of different lengths.}
    \label{fig:piv_supplemental}
\end{figure}

In Figure ~\ref{fig:V2-4mm} we show two additional kymographs for experiments with $L=2$ mm and $L=4$ mm respectively. Moreover, to examine more closely the velocity of the cells close to the fiber, we show in figure \ref{fig:piv_supplemental}b the vertically-averaged streamwise (x) velocity for three fibers of lengths $L = 2$, 3 and 4 mm. The vertical average has been done in blue bands of width $\Delta x = 500$ $\mu$m like the one illustrated in figure \ref{fig:piv_supplemental}a for a $L = 2$ mm fiber. This figure shows the goodness of the hypothesis of uniform velocity. Although this $x-$averaged velocity varies along the fiber length, $y$ direction, these variations (whose peak amplitude is about $\pm 50$\% of the average taken over the fiber's length) have a length scale shorter than the fiber, about 1 mm or less. Thus, it is reasonable as an approximation to assume that these fluctuations do not affect much the fiber deflection.
\\ 

\subsection*{Summary of force and fiber deflection data from experiments}
$L$ = Length of the fiber; $f_{0, max}$ = maximum force per unit length, corresponding to the maximum deflection; $x_{tip, max}$ = maximum deflection of the fiber tip; $x_{tip, end}$ = final displacement of the tip at the end of the experiment (120 hours).\\

\begin{center}
\begin{tabular}{|c|c|c|c|}
    \hline
     $L$ ($\mu m$) & $f_{0, max}$ ($nN/\mu m$)  & $x_{tip, max}$ ($\mu m$) &  $x_{tip, end}$ ($\mu m$)\\
     \hline
     \hline
     2006 & 10.94 & 456  &  343 \\ 
     1945 &  9.21 & 369  &  162 \\
     1984 &  6.98 & 271  &  168 \\
     2063 &  8.61 & 403  &  264 \\
     3018 &  3.92 & 837  &  591 \\
     3068 &  3.01 & 705  &  552 \\
     3007 &  4.25 & 904  &  796 \\
     2951 &  4.06 & 770  &  674 \\
     3928 &  2.31 & 1446 & 1345 \\
     4049 &  2.10 & 1419 & 1302 \\
     4096 &  2.28 & 1728 & 1630 \\
     4982 &  0.76 & 992  &  921 \\
     4988 &  0.60 & 675  &  - \\
     4957 &  1.76 & 1204 &  - \\
 \hline
\end{tabular}
\end{center}

Images from two of the experiments for the longest fibers ($L = 5$ mm) were not analyzed over time. We only fitted equation (\ref{eq:rod_shape}) to the last deformation of the fiber. For this length, the final fiber shape is very similar to that corresponding to the maximum deformation.\\

\subsection*{Cell density upstream and downstream the fiber}

\begin{figure}[hbt!]
\centering

\includegraphics[width=0.65\columnwidth,angle=270]{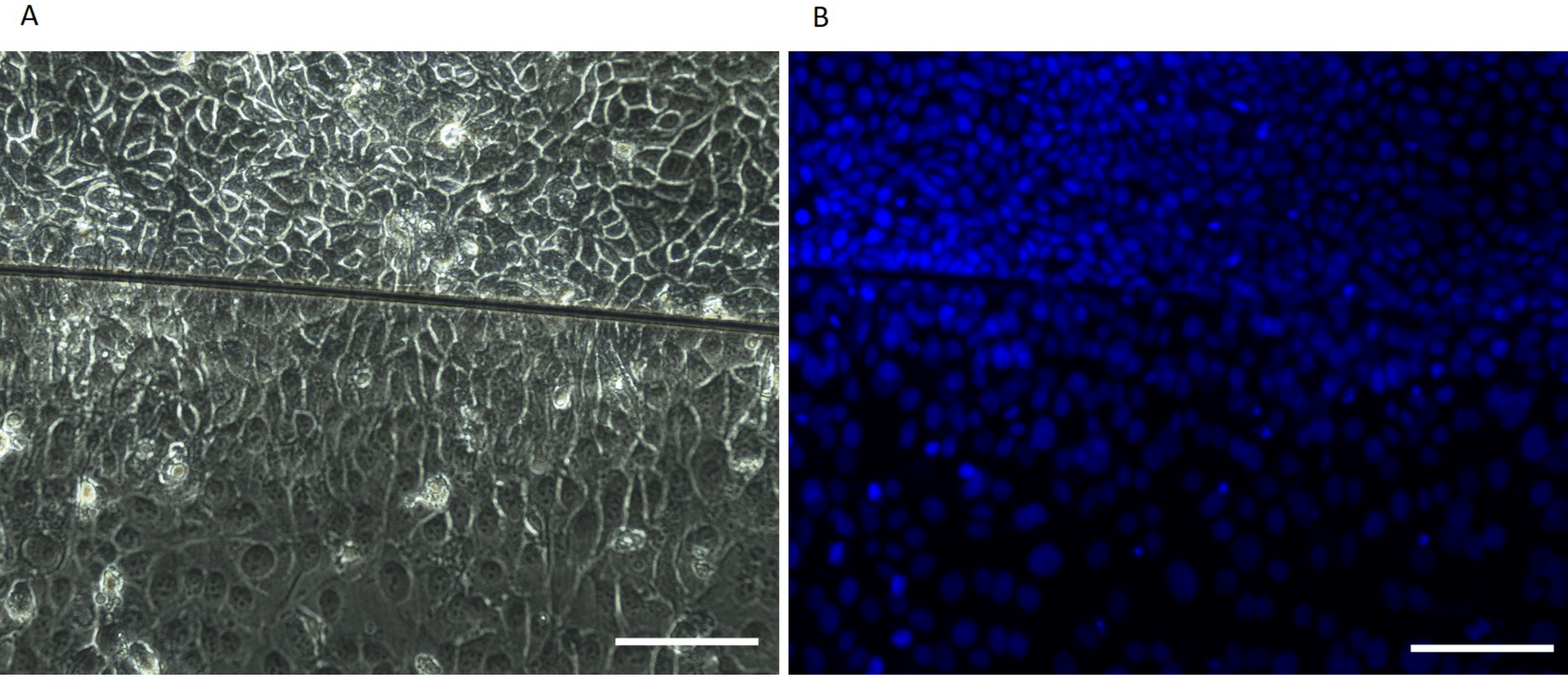} 
\caption{\label{fig:dapi} Migrating monolayer fixed at time = 48 hours. A Bright-field image. B. Nuclear staining with fluorescent DAPI (blue) of the same field shown in A. Carbon fiber is visualized as a black line. Scale bar: 100$\mu$m. For DAPI staining, cells are fixed with 4 \% paraformaldehyde in Phosphate-buffered saline (PBS) at room temperature for 20 minutes directly on the culture plate.
Then, they are incubated with DAPI at 1 $\mu$g/ml in PBS at room temperature for 5 minutes in the dark. The sample is washed with PBS and observed under the fluorescence microscope at Excitation/Emission of 358/461 nm.
}
\end{figure}

\begin{figure}[hbt!]
\centering
\includegraphics[width=0.6\columnwidth]{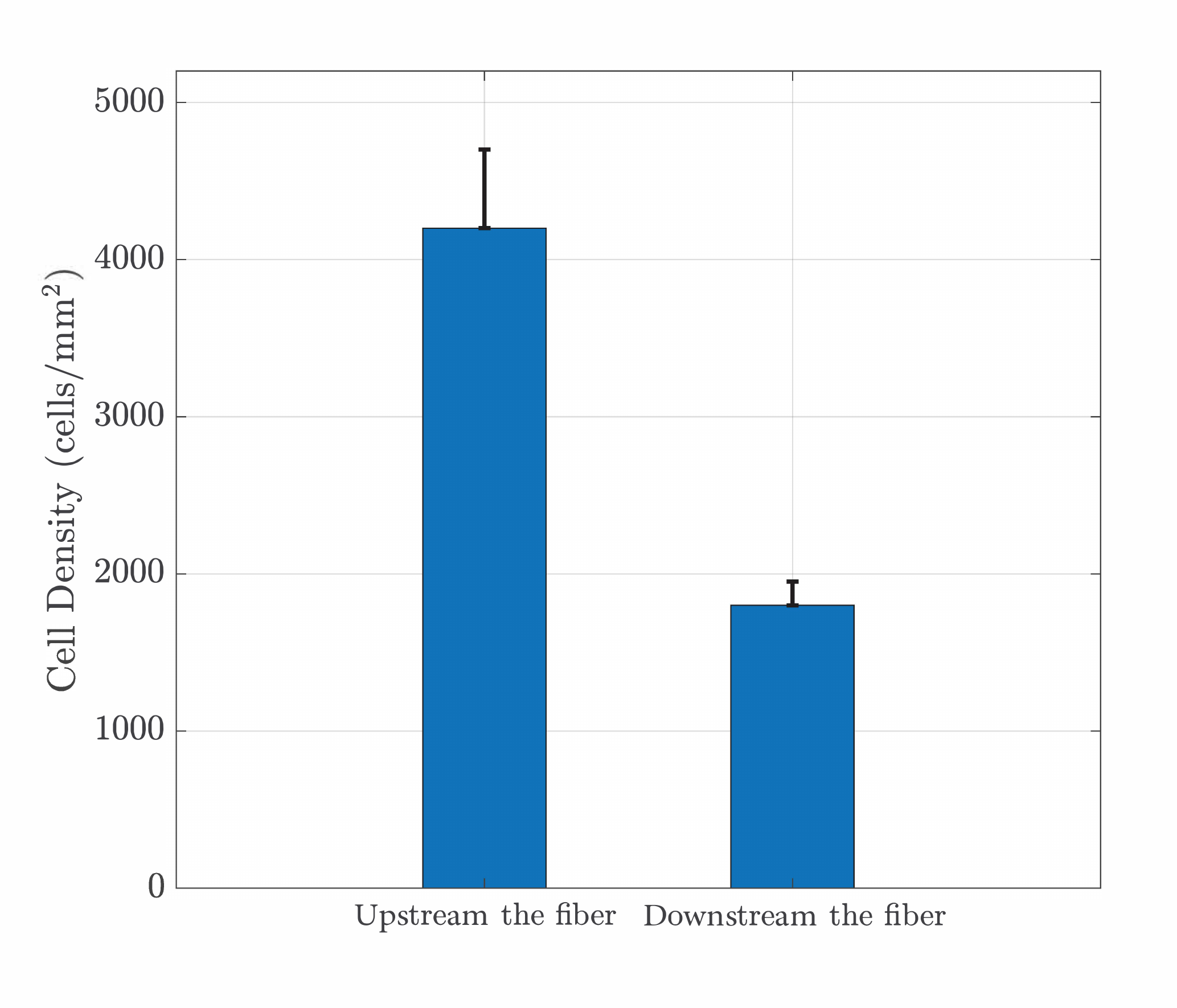}
\caption{\label{fig:cell_density} Cell density (Number of cells/ mm$^2$) analysis of Figure ~\ref{fig:dapi}.}
\end{figure}

Fig. ~\ref{fig:dapi} shows the  cell distribution upstream and downstream the fiber. Both bright-field images (Fig. ~\ref{fig:dapi}A) and nuclear staining (~\ref{fig:dapi}B) show that the cell density is much higher upstream the fiber, where cells can be seen to clump. The stretching of the cells that have migrated downstream the fiber is quite clear. Figure ~ \ref{fig:cell_density} shows a quantitative analysis of the DAPI image, to make more clear the very different densities found at both sides of the fiber.


\end{document}